\documentstyle[12pt,a4,epsf]{article}
\newcommand{\rcite}[1]{{\cite{#1}}}
\newcommand{\rref}[1]{{(\ref{#1})}}
\newcommand{\tref}[1]{{\ref{#1}}}
\newcommand{\rlabel}[1]{{\label{#1}}}
\newcommand{\rbibitem}[1]{\bibitem{#1}}
\newcommand{\be}{\begin{equation}}
\newcommand{\ee}{\end{equation}}
\newcommand{\ba}{\begin{eqnarray}}
\newcommand{\ea}{\end{eqnarray}}
\newcommand{\dis}{\displaystyle}

\newcommand{\mathrm}[1]{{\rm #1}}

\begin{document}
\begin{titlepage}
\begin{flushright}
{UG-FT-75/97}\\
{hep-ph/9708395}\\
\end{flushright}
\vspace{2cm}
\begin{center}
{\large\bf $m_u + m_d$ from QCD-Hadron Duality\footnote{Work supported
in part by CICYT Grant No. AEN-96/1672 (Spain).
Invited talk at ``High Energy Conference on Quantum Chromodynamics
 (QCD '97)'', 3-9 July 1997, Montpellier, France.}}\\
\vfill
{\bf Joaquim Prades}\\[0.5cm]
Departamento de
 F\'{\i}sica Te\'orica y del Cosmos, Universidad de Granada\\
Campus de Fuente Nueva, E-18002 Granada, Spain.\\[0.5cm]
\end{center}
\vfill
\begin{abstract}
 The value of the light quark masses combination $m_u + m_d$
is analized using QCD-Hadron Duality. A detailed analysis of both the
perturbative QCD [to four-loops] and the hadronic parametrization 
needed is done. The result we get is $[m_u + m_d] (1 \, {\rm GeV}^2) =
(12.8 \pm 2.5)$ MeV ($[m_u + m_d] (4 \, {\rm GeV}^2) = 
(9.8 \pm 1.9)$ MeV ) in the $\overline{\rm MS}$ scheme.
\end{abstract}
\vfill
August 1997
\end{titlepage}

We are interested in the value of the light quark masses
that appear in the Standard Model. These are the QCD Lagrangian
quark masses or Current Algebra quark masses. 
This work is an  update of \rcite{BPR95} putting some emphasis in the
QCD-Hadron Duality and including the recently calculated $\alpha_s^3$
QCD corrections \rcite{C97}. A collection  of the relevant references
to earlier work can be found in \rcite{BPR95}. 

The QCD Sum Rule (SR) technique has been discussed several times in this
conference. (See the talks by Narison and Jamin for instance.) 
The relevant two-point function for calculating $m_u+m_d$ is
the correlator of two divergences of the axial-vector current with 
the quantum
numbers of the pion field $\left[ I^G (J^P) = 1^- (0^-) \right]$.
(See also the talk by de Rafael.)
\ba
\Psi_5(q^2)\equiv i {\dis \int} {\rm d}^4 x \, e^{iq\cdot x}
\, \langle 0 | T \left\{ \partial^\mu A_\mu^{1-i2}(x),  
\partial^\nu A_\nu^{1+i2}(0) \right\} | 0 \rangle \, ,  
\ea
with $ \partial^\mu A_\mu^{1-i2}(x)= \left[ 
\partial^\mu A_\mu^{1+i2}(x) \right]^\dagger = 
\left[ m_u + m_d \right]
\, \left[ \overline u  i \gamma_5 d \right]$. 
The QCD Sum Rule technique exploits the  analytic structure  of 
 $\Psi_5(q^2)$ combined with Cauchy's theorem.
The specific type of SR we shall use
is the so-called Finite Energy Sum Rule (FESR). 
For the discussion of quark masses using QCD-Hadron
Duality there are two relevant FESRs. These are the  so-called 
0th-moment
\ba
\rlabel{0mom}
{\dis \int^s_0} {\rm d} t \, \frac{\dis 1}{\dis \pi}
\, {\rm Im} \Psi_5(t)  &=& \frac{\dis N_c}{\dis 8 \pi^2} \, 
\left[ m_u(s)+m_d(s)\right]^2 \, \frac{\dis s^2}{\dis 2} \,
\left[ 1 + R_1(s) + 2 \, \frac{\dis C_4 \langle O_4 \rangle}{\dis
s^2}  \right] \nonumber \\ 
 \ea
and  1st-moment
\ba
\rlabel{1mom}
{\dis \int^s_0} {\rm d} t \, t \, \frac{\dis 1}{\dis \pi}
\, {\rm Im} \Psi_5(t)  &=& \frac{\dis N_c}{\dis 8 \pi^2} \, 
\left[ m_u(s)+m_d(s)\right]^2 \, \frac{\dis s^3}{\dis 3} \,
\left[ 1 + R_2(s) - \frac{\dis 3}{\dis 2} \, 
\frac{\dis C_6 \langle O_6 \rangle}{\dis s^3}  \right] \, ,
\nonumber \\
 \ea
respectively. The upper limit of the integral, $s$,  is the onset
of the QCD continuum. The hypothesis of QCD-Hadron Duality that we have 
used to obtain \rref{0mom} and \rref{1mom}
can be expressed as follows. One expects that there exists 
some intermediate region [$s$ belongs to that region], such that
above this region   the description of inclusive enough quantities like
$\Psi_5(q^2)$ are well approximated by perturbative QCD, i.e. local duality
is approximately satisfied. In the intermediate region
one can just expect to have duality with perturbative QCD for
suitable averages of inclusive quantities like the moments
in \rref{0mom} and \rref{1mom}, i.e. global duality.
 Below that region one doesn't expect any kind of duality.

Let me discuss the inputs needed in these FESRs. The 
rhs in \rref{0mom} and  \rref{1mom} are given by 
\ba
\rlabel{cau}
-\frac{\dis 1}{\dis 2 \pi i} \, 
{\dis \oint_s} {\rm d} t \, \Psi_5^{(2)}(t)  \, 
  \left(1-\frac{\dis t}{\dis s} \right)^2 \, 
{\dis \sum_{k=1}^n} \frac{\dis n+1-k}{\dis n}
 \, \left(\frac{\dis  t}{\dis s}\right)^{k-1}  
\ea
with $n=1$ and $n=2$ respectively, and $\Psi_5^{(2)}(t)
\equiv \frac{{\rm  d}^2}{{\rm d} t^2} \, \Psi_5(t)$.
 For large enough $Q^2 \equiv -q^2 = -t$, $\Psi_5^{(2)}(Q^2)$ is known 
in QCD  up to four-loops [order $\alpha_s^3$] from \rcite{C97,RUS}. 
Once the global light quark masses factor
scaled at $s$ together with the energy dependence and normalization
factors have been pulled out, what remain from \rref{cau} is the
 QCD perturbative series $1 + R_i(s)$.
This is supplemented by  power corrections {\em \`a la}
SVZ \rcite{SVZ79} parametrized by Wilson coefficients
 $C_i$ and condensates $\langle O_i \rangle$.  
The values of the power corrections coefficients are
$C_4 \, \langle O_4 \rangle = (0.08 \pm 0.04)$ GeV$^4$ and 
$C_6 \, \langle O_6 \rangle = (0.04 \pm 0.03)$ GeV$^6$ \rcite{BPR95}.

To study the convergence of the  truncated perturbative series,
 we have used three different ways of resumming the QCD series.
The first one, that we shall call {\em Perturbative}, is 
obtained resumming the renormalization group logs {\em after}
the Cauchy's integral in \rref{cau} is done. In the second one,
which we shall call {\em Improved},  the large
 running of the coupling constant around the Cauchy's circuit 
(from 0 to 2$\pi$) is taken into account by resumming the 
renormalization group logs {\em before}
the Cauchy's integral in \rref{cau} is done \rcite{LP92}. 
The third way of resumming is obtained by applying the Principle of Minimal
Sensitivity  ({\em PMS}) \rcite{ST81} to the  QCD expression 
for $\Psi^{(2)}(Q^2)$ {\em before} the Cauchy's integral is done.
We have applied the renormalization group running 
to $\alpha_s(m_\tau) = 0.35 \pm 0.02$  \rcite{PI97} at  three-loops to obtain 
$\alpha_s(s)$ around the Cauchy's circuit. 
Below we show the QCD perturbative series $1+ R_2(s)$
for $s=$ 2 GeV$^2$ obtained using the three ways of resumming.
\ba
\rlabel{resum}
Perturbative &:& 1 + 0.60 + 0.38 + 0.19 + \cdots \nonumber \\
Improved &:& 1 + 0.68 + 0.20 + 0.07 + \cdots \nonumber \\
PMS &:& 1 + 0.85 + 0.12 - 0.01 + \cdots \, .
\ea
The second term in each line is the order $\alpha_s$ correction,
 the third is the order $\alpha_s^2$ and the fourth the order
$\alpha_s^3$ correction. The comparison of the series in \rref{resum}
shows that {\em PMS} is the best behaved but also shows
how the three series show good convergence {\em if}
one considers corrections up to order $\alpha_s^3$.
The numerical difference between them  is less than 10 \% at that order.

Let me now discuss the inputs needed for the lhs of 
\rref{0mom} and \rref{1mom}. The integrand ${\rm Im} \Psi_5(t)$ 
is an inclusive cross-section which, ideally, should be 
determined in terms of the hadronic states that couple to the vacuum 
through the divergence of the axial-vector current 
$ \partial^\mu A^{1-i2}_\mu(x)$, i.e. the $\pi$-pole, 3$\pi$-continuum, 
5$\pi$-continuum, $\pi'$-resonances, $N \overline N$, $\cdots$.
Apart of the pion pole, we have not complete access to this 
spectral function yet. We do however have several pieces of information 
on it.  In the low energy end, Chiral Perturbation Theory (CHPT) predicts 
\ba
\rlabel{spectral}
\frac{\dis 1}{\dis \pi} {\rm Im} \Psi_5(t)
&=& 2 f_\pi^2 m_\pi^4 \, \delta (t- m_\pi^2) \nonumber
\\ &+&  
\Theta (t- 9 m_\pi^2) \, \frac{\dis 2 f_\pi^2 m_\pi^4}{\dis
\left[ 16 \pi^2 f_\pi^2 \right]^2} \, \frac{\dis t}{\dis 18} \, 
 \rho^{3\pi}_\chi(t) \,+ \, \cdots
\ea
where $f_\pi=$ 92.4 MeV in this normalization and the dots stand
for higher thresholds contributions. The first term is due to the pion
 pole and is completely 
fixed by the pion mass and its decay coupling constant $f_\pi$,
this will become important afterwards.
When $t \to 0$, the spectral function $\rho^{3\pi}_\chi(t)$
can be obtained at lowest order in CHPT. At lowest order in 
Generalized CHPT, one gets the CHPT result times some factor, 
see Stern's talk. In fact, no
 symmetry allows to predict $\rho^{3\pi}_\chi(t=0)$ and 
it receives corrections to all orders in CHPT proportional 
to pion  and kaon masses. We shall let this normalization factor free
and require QCD-Hadron Duality to fix it, see below.

For values of $t$ above a few hundred MeVs, 
CHPT is not anymore at work and we enter the resonance region. 
Here we  have some  experimental information to help us, for instance
we know that there are resonance states with the quantum numbers of 
the pion which couple to the 3$\pi$-continuum, 5$\pi$-continuum,
$\cdots$. There are also two-body thresholds like  $\rho \pi$, 
$\rho' \pi$, $\rho \pi'$, $N \overline N$, and so on. 
The contributions of the resonances  
are in fact leading in the $1/N_c$ expansion 
 \rcite{BPR95,PR97} and have to be included if we want to
obtain QCD-Hadron Duality. 
In the Review of Particle Physics \rcite{PDG96}, we can find two 
resonances with the pion quantum numbers. Their masses and widths are: 
$M_1 = [1300 \pm 100]$ MeV, $\Gamma_1 = [400 \pm 200]$  MeV,
$M_2 = [1770 \pm 30]$ MeV,  $\Gamma_2 = [310 \pm 50]$ MeV.
The most recent experimental
data where both $\pi'$-resonances are seen is in \rcite{BE84}. 
One can find  there the
energy dependence of the $I^G(J^P)=1^-(0^-)$ amplitude
in the three-pion channel.  
This amplitude is proportional [in the three-pion channel] to the 
spectral function with the quantum numbers  of
the divergence of the axial-vector current we are studying.
Unfortunately, the normalization of the coupling of this current 
to $N \overline N$ is unknown.
That data allowed to obtain the masses and widths of the 
two $\pi'$-resonances since the non-resonant background
in the data was found to be  very small \rcite{BE84}.
The values obtained there agree quite well with the ones reported in 
the Review of Particle Physics \rcite{PDG96}.
Therefore, we shall take the $I^G(J^P)=1^-(0^-)$ data in 
\rcite{BE84} as a good estimate of the
resonant shape allowing a free relative strength between both
$\pi'$ resonances. 

Another piece of information to add to our study
of the spectral function is the success of the resonance dominance
 in the vector channel (VMD). One can obtain the successful VMD 
predictions using  a spectral function with 
Breit-Wigner shapes for the resonances normalized to  
lowest order CHPT prediction\rcite{BPR95,PR97}.
The difference in this case with respect to the
divergence of the axial-vector channel one
 is that the normalization at $t=0$ of the vector form factor is fixed 
since the vector current couples to the electromagnetic current.
The proposal we make is to use scalar meson dominance for the
$I^G(J^P)=1^-(0^-)$ channel, so that the spectral function 
for the three-pion channel is the one in \rref{spectral} with
the substitution 
\ba
\hat \rho_\chi^{3\pi} (t) 
\to  \rho^{3\pi} (t) \equiv {\cal A} \, \hat \rho_\chi^{3\pi} (t) 
\, \left| {\cal F} (M_1, \Gamma_1, M_2, \Gamma_2; t) \right|^2
\ea
where ${\cal A}$ is a free normalization factor, 
$\hat \rho_\chi^{3\pi} (t)$ is the lowest order
CHPT result and
\ba
{\cal F} (M_1, \Gamma_1, M_2, \Gamma_2; t)
\equiv \frac{\dis 
\frac{\dis 1}{\dis t-M_1^2 + i \Gamma_1 M_1} 
+ \xi \frac{\dis 1}{\dis t-M_2^2 + i \Gamma_2 M_2} }
{\dis  \frac{\dis 1}{\dis -M_1^2 + i \Gamma_1 M_1} 
+ \xi \frac{\dis 1}{\dis -M_2^2 + i \Gamma_2 M_2}  } \, .
\ea
Here $\xi$ is a complex parameter which fixes the
relative strength between the two $\pi'$ resonances
and a possible different absorption by the nuclear target.
We also included the corresponding Breit-Wigner resonant shape
for the $\rho$-meson in the intermediate $\rho \pi$ 
two-body subchannel which contribution is leading in $1/N_c$
and phase space \rcite{BPR95,PR97}.
The $\xi$ parameter can be obtained from a fit of the
$I^G(J^P)=1^-(0^-)$ data in \rcite{BE84}.
We get a good fit to the  shape of the 
$I^G(J^P)=1^-(0^-)$ data by fixing $\xi = 0.234 + i \, 0.1$.

Let me now apply the technique to obtain $m_u+m_d$
 using all the information above. We only  use the
three-pion channel. The onset of the QCD continuum $s$ in \rref{0mom}
and  \rref{1mom} is fixed by demanding a good QCD-Hadron 
Duality between
\ba
{\cal R}_{\rm had} (s) = \frac{\dis 3}{\dis 2 s} \, 
\frac{\dis {\dis \int^s_0} {\rm d} t \, t \,
\frac{\dis 1}{\dis \pi} {\rm Im} \Psi_5(t)}
{\dis {\dis \int^s_0} {\rm d} t \, 
\frac{\dis 1}{\dis \pi} {\rm Im} \Psi_5(t)}
& {\rm and} & \rlabel{qcdcount}
{\cal R}_{\rm QCD} (s) = \frac{\dis 1 + R_2(s) 
- \frac{\dis 3}{\dis 2} \frac{\dis C_6 \langle
O_6 \rangle}{\dis s^3}}{\dis 1 + R_1(s) 
+ 2 \frac{\dis C_4 \langle O_4 \rangle}{\dis s^2}}\, .
\nonumber \\
\ea
The plots of the ratios in 
\rref{qcdcount} are shown in Figure \tref{duality}. 
\begin{figure}[htb]
\begin{center}
\leavevmode\epsfxsize=9.3cm\epsfbox{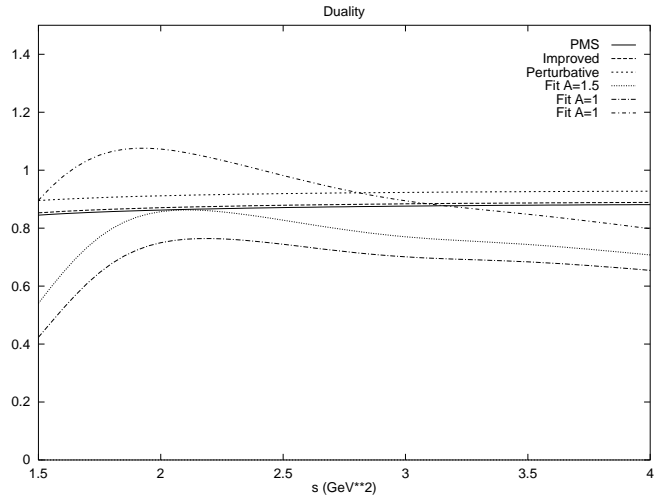}
\caption{\rlabel{duality} Duality Ratios in 
\protect{\rref{qcdcount}}. See text  for explanation of the curves.}
\end{center}
\end{figure}
The curves labeled as {\em Perturbative}, {\em Improved},
and {\em PMS} correspond to ${\cal R}_{\rm QCD}(s)$
obtained using the different ways of resumming the QCD series
explained above. Observe the small difference between the three curves.
The effect of varying $\alpha_s$ and the condensates in the range 
quoted above is less than 5 \%. 
 Other non-perturbative contributions like for instance
the possible large contribution of instantons 
mostly cancel in ${\cal R}_{\rm QCD}(s)$. In particular, the 
result in \rcite{GN93} produces less than 5 \% change in 
${\cal R}_{\rm QCD}(s)$. After the good control we have on
the QCD counterpart, a violation of global QCD-Hadron
Duality larger than say 10 \% is unexpected and certainly would 
have to be understood. 

Notice that duality {\em cannot} diiferentiate between shapes
of spectral functions which have the area below.
 On the contrary, the interference of the pion pole 
contribution in \rref{spectral}
and the three-pion contribution makes the ratio ${\cal R}_{\rm had}(s)$
sensitive to changing the global normalization factor ${\cal A}$.
The normalization factor ${\cal A}$ should be around 1.5
if QCD-Hadron Duality is wanted (We have used the {\em PMS}
 resummed series as the best behaved.). 
In Figure \tref{duality} we show three different normalizations:
namely, the lower curve is for ${\cal A}=1$ in analogy with the 
normalization of the conserved vector current; the curve  
with  normalization factor ${\cal A}=$ 1.5 is the one that
has the best duality with the {\em PMS} resummed QCD series,
 and the higher curve is for  ${\cal A}=5$ where more than
20 \% violation of duality is obtained.  Notice also that one cannot expect
duality for $s$ larger than [3 $\sim$ 3.5] GeV$^2$ since from these
energies on there are two-body thresholds 
like $N \overline N$, $\rho \pi'$ [which couples dominantly 
to 5$\pi$-continuum], $\cdots$. None of these intermediate states have been
included in our parametrization or in the one in \rcite{BGM97}.
These contributions are leading in the $1/N_c$ counting and dominant by
phase space. 
This was overlooked in \rcite{BGM97} where duality in 
the three-$\pi$ channel was required at energies much higher 
than 4 GeV$^2$. More on this issue in \rcite{PR97}.

Once we have fixed the duality region from Figure
\tref{duality} to be around 2 GeV$^2$ we can resolve for the
light quark masses in that region. Using the 0-th moment \rref{0mom},
we get  the masses in Figure  \tref{masses}.
\begin{figure}
\begin{center}
\leavevmode\epsfxsize=9.3cm\epsfbox{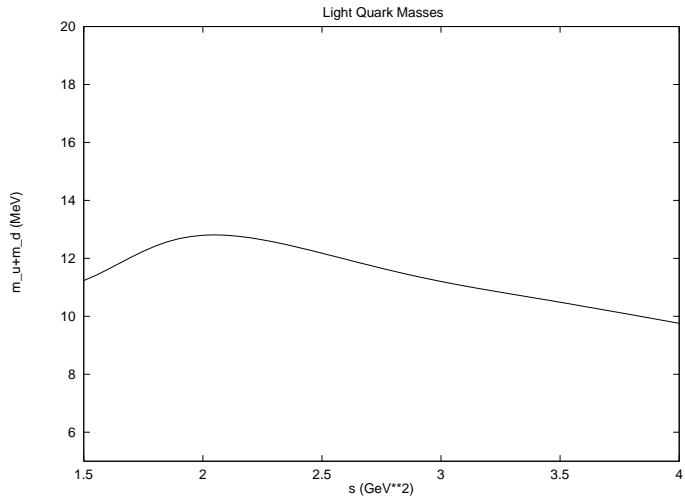}
\end{center}
\caption{\rlabel{masses} $m_u+m_d$ $\overline{\rm MS}$ masses
scaled at 1 GeV.}
\end{figure}
There we plot the $\overline{\rm MS}$ $m_u + m_d$ masses
scaled to 1 GeV  obtained using the hadronic parametrization
with ${\cal A} =$ 1.5 and the {\em PMS} resummed QCD series.
The value of $m_u+m_d$ in the duality region can be read off
from Figure \tref{masses}. The result we get is 
\ba
\rlabel{mass1}
[m_u + m_d] (1 {\rm GeV}^2) = 
( 12.8 \pm 2.5) \quad {\rm MeV}
\ea
i.e.
\ba
\rlabel{mass4}
[m_u + m_d ](4 {\rm GeV}^2) = 
( 9.8 \pm 1.9) \quad {\rm MeV}
\ea
in the $\overline{\rm MS}$ scheme at order $\alpha_s^3$.
This result updates the one in \rcite{BPR95}.
The central value is for ${\cal A}=$ 1.5 and using the {\em PMS} 
resummed QCD series. The error includes the hadronic uncertainty
allowing for a violation of duality of more than 10 \% at $s=2$
GeV$^2$ and by varying $\alpha_s$ and the coefficients of the
power corrections in the ranges given. All these uncertainties have 
been added in quadrature. Notice also that the results one gets
using the {\em Improved} QCD resummed series or the {\em PMS}
one only differ by 0.1 MeV. This shows the good control we have on
the QCD series. Even though our spectral function is obviously not
complete [see comments above], one can see from Figure \tref{masses}
that the value of the masses is quite stable
between $s=2$ GeV$^2$ and $s=4$ GeV$^2$. 
Instanton contributions to the 0-th moment for $s$ around 3 GeV$^2$
are negligible \rcite{GN93} and have been included in the quoted error.
Laplace Sum Rules give compatible results.
The results in \rref{mass1} and \rref{mass4} are in perfect
agreement with the bounds obtained recently in \rcite{LRT97}.

Given the good behaviour shown by the QCD counterpart, further
improvements reducing the error bars in the determination of
$m_u+m_d$ have  to come from the very difficult measurement
of the $I^G(J^P)=1^-(0^-)$ spectral function normalization. 
As we tried
to show, our present understanding already gives tight bounds on the
values of $m_u+m_d$.

I would  like to thank Hans Bijnens and Eduardo de Rafael
for reading the manuscript and useful comments.

\end{document}